\documentclass[12pt]{article}

\usepackage{amssymb,amsmath} 
\usepackage{amsfonts}
\usepackage[body={17cm,23.5cm}]{geometry}
\begin{document}

\title{\bf  Physics and Cosmology : \\ the Milli-Electron-Volt Scale\footnote{Talk given 
at the {\it IV Symposium on
large TPCs for low-energy rare event detection}, Paris, December 2008.} 
}
\author{ \bf{Eduard Mass{\'o}}\\
\small{ \it Grup de F{\'\i}sica Te{\`o}rica and Institut de F{\'\i}sica d'Altes Energies}\\
\small{  \it Universitat Aut{\`o}noma de Barcelona, 08193 Bellaterra, Spain}
}

\date{\small{Preprint UAB-FT-662}}

\maketitle

%\vspace*{.5cm}
\centerline{ \large \bf Abstract}
%This is the written version of a talk at the {\it IV Symposium on
%large TPCs for low-energy rare event detection}, held in Paris (December 2008).
A short review about  vacuum energy and the cosmological constant is presented.  
The observed acceleration of the universe introduces a new meV energy scale. The problem is that, theoretically, the predicted vacuum energy is many orders of magnitude larger than $10^{-3}$ eV.
The problem is a link between two Standard Models, namely the Standard Model of Particles and their Interactions (where the vacuum energy appears) and the Standard Cosmological Model (where a cosmological constant is a good fit to data), and perhaps it is a clue in our search for new physics.

%\vspace*{0.5cm}
 \begin{center}
---------------------------------------------------------------------
 \end{center}
 
%\vspace*{0.5cm}

%\label{sec:1}
% Always give a unique label
% and use \ref{<label>} for cross-references
% and \cite{<label>} for bibliographic references
% use \sectionmark{}
% to alter or adjust the section heading in the running head

\section{The Cosmological Constant}

Let us mention some important facts about our universe. First of all, we know our universe is expanding because of the measured redshift in the spectral lines of light sent by distant objects. Consider the moment of emission of a photon from a galaxy which is at a (physical) distance $d$ from the Earth. The photon is emitted with a (proper) wavelength
 $\lambda_e$. At the moment of reception of the photon at the Earth the physical distance between the galaxy and us has increased to $d_0$ and the wavelength  $\lambda_r$ of the photon has increased in the same proportion
 \begin{equation}
\label{ }
\frac{\lambda_r}{\lambda_e} = \frac{d_0}{d} = 1 + z
\end{equation}
where $z$ is the redshift.

Introducing the assumption of homogeneity and isotropy, we can describe our universe quite simply
\cite{kt}. Before showing it, we introduce the concept of comoving distance and scale factor. Take the example of the galaxy we 
introduced before and the Earth. Their comoving distance $x$ is fixed and tells us which two objects we are considering. The expansion can be described in terms of a time dependent scale factor $a(t)$, in such a way that at the moment of emission the scale factor has a value $a=d/x$ while at reception it is $a_0=d_0/x$. 

The equations governing the behavior of the scale factor can be deduced from the Einstein equations of General Relativity,
\begin{equation}
\label{EinsteinEquations}
R_{\mu\nu} - {1\over2}g_{\mu\nu}R + \Lambda \,g_{\mu\nu}= 8\pi G_N\,T_{\mu\nu}
\end{equation}
where $T_{\mu\nu}$ is the energy-momentum tensor of the universe content and
the parameter $\Lambda$ is the (in)famous cosmological constant. Let us first set $\Lambda=0$. In this case, from the Einstein equations we obtain the Friedmann equations
\begin{equation}
\label{friedmann}
H^2 = \left({\dot a\over a}\right)^2 = {8\pi G_N\over3}\,\rho 
\end{equation}
\begin{equation}
\label{2a}
{\ddot a\over a} = -\,{4\pi G_N\over3}\,(\rho + 3p)
\end{equation}
Here $\rho$ and $p$ include the energy density and pressure of the different fluids that form the universe. We have simplified a bit the equations because we have written them in the case of a flat universe, which is a good approximation.

Observations at recent $z$ favor a universe with $\ddot a >0$, i.e.,  an accelerating universe \cite{Frieman:2008sn}. We know from independent observations that there is a large density of cold dark matter in the universe. However its contribution to (\ref{2a}) would lead
to $\ddot a <0$ since $\rho_{ m} >0$ and $p_{m} =0$ (``\textit{m}"  means both visible matter and dark matter).
Actually, fits to the data lead to $p$ in (\ref{2a}) to be negative and on the order of magnitude of $p \sim - \rho$. From where we may have negative pressure contributions?

The cosmological constant may come to our rescue. The full Friedmann equations when $\Lambda$ is present are
\begin{equation}
\label{ }
H^2 = \left({\dot a\over a}\right)^2 = {8\pi G_N\over3}\,\rho +
{\Lambda\over3}
\end{equation}
\begin{equation}
\label{2a_new}
{\ddot a\over a} = -\,{4\pi G_N\over3}\,(\rho + 3p) + {\Lambda\over3}
\end{equation}
We now see that the $\Lambda$ contribution is able to give an accelerating universe, 
$\ddot a >0$.

It is convenient to write an equivalent density $\rho_\Lambda$ and pressure $p_\Lambda$ for the cosmological constant contribution,
\begin{equation}
\label{ } {\Lambda\over3 }
 \equiv {4\pi G_N\over3}\rho_\Lambda
\end{equation}
\begin{equation}
\label{ }
p_\Lambda=-\rho_\Lambda
\end{equation} 
With these definitions the form of the Friedmann equations is simply given by 
(\ref{friedmann}) and (\ref{2a}), but now understanding that there are contributions not
only from visible matter and dark matter but also from $\Lambda$,
\begin{eqnarray}
\label{ }
\rho &=&  \rho_{m}  + \rho_\Lambda + \dots \nonumber \\
p &=&  p_{m} + p_\Lambda + \dots
\end{eqnarray}
where the dots are potential components of the universe apart from matter and $\Lambda$.
(We neglect radiation because it does not play any role in the Friedmann equations in the universe at recent $z$).

It is worth noticing that these two components have a very simple equation of state relating density and pressure,
\begin{equation}
\label{ }
p_i = w_i \rho_i
\end{equation}
with $w_i$ constant. Indeed, we have
\begin{eqnarray}
w_{m} & = &  0 \nonumber \\
w_{\Lambda} & = & -1
\end{eqnarray}
In these cases one can integrate the equation
\begin{equation}
\label{ }
{d \over dt}\, (\rho_i a^3) + p_i \,  {d \over dt}\, (a^3)=0
\end{equation}
which can be deduced from (\ref{friedmann}) and (\ref{2a}), and which expresses energy conservation. One gets for matter
\begin{equation}
\label{ }
\rho_{ m}  \propto \  {1\over a^3}
\end{equation}
expressing the fact that volume goes as $a^3$ in an expanding universe. However, for the cosmological constant one gets
\begin{equation}
\label{ }
\rho_{\Lambda}  \propto \  {\rm constant}
\end{equation}

\section{Dark Energy}

To obtain the values for the matter and the cosmological constant densities, one uses data from supernovae, cosmic microwave background, cluster analysis, large scale structure, age, lensing, etc.
What one gets is a consistent and robust picture, called the Standard Cosmological Model where the total density of universe is very near the critical density $\rho_c$ (defined as the density that would make the universe exactly flat),
\begin{equation}
\label{ }
\rho_m + \rho_\Lambda \simeq \rho_c
\end{equation}
and where the contributions are
\begin{equation}
\label{ }
\Omega_{m}={\rho_{m} \over \rho_c} \simeq 0.3
\end{equation}
and
\begin{equation}
\label{ }
\Omega_{\Lambda}={\rho_{\Lambda} \over \rho_c} \simeq 0.7
\end{equation}
with small errors.

This last contribution can be expressed back in terms of $\rho_\Lambda$, and one
gets
\begin{equation}
\label{ }
\rho_\Lambda \simeq (2 \times 10^{-3}\ {\rm eV})^4
\end{equation}
This is the new meV energy scale that characterizes the cosmological contribution.
To understand its origin is a challenge for contemporary physics.

We have used the contribution of the cosmological constant to describe the acceleration of the universe, but it is time now to make some generalization. We could have contributions in (\ref{2a}) that lead to acceleration without the need to be exactly of the form of a cosmological constant.
We could have for instance $w \neq -1$, but still leading to acceleration. We see from (\ref{2a}) that $w< -1/3$ gives $\ddot a >0$. In all generality we could have a $z$-dependent $w=w(z)$.
We call dark energy any component of the universe leading to acceleration. We have in this case a density  $\rho_{de}$ that today is measured as
\begin{equation}
\label{de_exp}
\rho_{de} \simeq (2 \times 10^{-3}\ {\rm eV})^4
\end{equation}

A popular model of dark energy is quintessence, where fields which are slowly varying in time are the origin of the acceleration \cite{Frieman:2008sn}. 

\section{Vacuum Energy}

The vacuum energy acts as a cosmological constant. This is due to the fact that general covariance gives the form of the expectation value in the vacuum of the energy-momentum tensor 
\cite{Zel'dovich:1968zz},
\begin{equation}
\label{ }
T_{vac}^{\mu \nu} =<0| T^{\mu \nu}  |0>  = \rho_{vac}\ g^{\mu \nu}
\end{equation}
This might be good news, because vacuum energy could be an explanation for having a cosmological constant $\Lambda$.

The problem arises when we realize that the Standard Model of Particles is a Quantum Field Theory, and that in vacuum the expectation value of the Hamiltonian does not vanish because it receives contributions of all zero-point energies of all modes for all fields
\begin{equation}
\label{divergence_H}
<0| H |0> = \sum_{fields} \eta_i \int_0^ \infty \ \frac{d^3 k}{(2\pi)^3}\  \frac{1}{2} 
\hbar \sqrt{k^2 +m_i^2}
\end{equation}
Here $\eta_{boson}=1$ and $\eta_{fermion}=-1$. The integral is divergent so that we obtain  infinite as the result.

What we do in Quantum Field Theory is to subtract this (infinite) quantity,
\begin{equation}
\label{ }
H \longrightarrow\ H - <0| H |0>
\end{equation}
or, in more technical terms, we use normal ordering. The argument is that we are only concerned with energy differences, so that the subtraction does not have any physical consequences.
However, this is true when we do not consider gravitational effects and it is no longer true when gravitation is on. 

The divergence in (\ref{divergence_H}) arises because we integrate over all energies and momenta, or, in other words, we believe the theory is valid up to infinite energies. Let us be more modest and assume we can use the Quantum Field theory until a maximum cutoff $k_{max}$. Thus for a single (bosonic) field we have
\begin{equation}
\label{ }
\rho_{vac} = \frac{1}{4 \pi^2} \int_0^{k_{max}} \ d k\ k^2 \sqrt{k^2 +m^2} = 
\frac{k^4_{max}}{16\pi^2}
\end{equation}
(we use $m \ll k_{max}$).

If we believe that our theories can be used until the Planck mass, then we get a total vacuum energy
\begin{equation}
\label{ }
\rho_{vac} =  \frac{M_P^4}{16\pi^2} \simeq (10^{18}\ {\rm GeV})^4
\end{equation}
We see that this estimate exceeds by many orders of magnitude the measured value (\ref{de_exp}).  Perhaps we are still too pretentious because in reality we have probed the Standard Model of Particles only until the Fermi scale $M_W$. If we use this cutoff we get
\begin{equation}
\label{ }
\rho_{vac} =  \frac{M_W^4}{16\pi^2} \simeq (10\ {\rm GeV})^4
\end{equation}
still 13 orders of magnitude in energy scale.

In conclusion, vacuum energy contributes to the energy density  
of the universe many orders of magnitude above observation.
Even if we find a theory that explains the acceleration of the universe with
the correct value (\ref{de_exp}), we should find
a satisfactory explanation to the problem of the vacuum energy contribution 
to the cosmological constant. Perhaps when trying to solve this challenging problem we will get hints  leading to the discovery of new physics \cite{Weinberg:1988cp}.

\section*{Acknowledgments}
 I am grateful to Ioanis Giomataris for the invitation to the workshop.
 I acknowledge support by the CICYT Research Project FPA 2008-01430 and the
\textit{Departament d'Universitats, Recerca i Societat de la
Informaci{\'o}} (DURSI), Project 2005SGR00916.
This work was supported (in part) by the European Union through the Marie
Curie Research and Training Network "UniverseNet" (MRTN-CT-2006-035863)."

\end{document}